\documentclass[epj]{svjour}
\usepackage{graphics,enumerate}
\usepackage{cite, hyperref}
\usepackage{ulem, color}
\usepackage{amsmath,amssymb, graphicx}
\usepackage[utf8]{inputenc}
\usepackage{color}

\def\CF{\mathrm{C_F}}
\def\CFsq{\mathrm{C_F^2}}
\def\CFcub{\mathrm{C_F^3}}
\def\CFfour{\mathrm{C_F^4}}

\def\CA{\mathrm{C_A}}
\def\CAsq{\mathrm{C_A^2}}
\def\CAcub{\mathrm{C_A^3}}

\def\W{\mathcal{W}}
\def\oW{\overline{\mathcal{W}}}

\def\A{\mathcal{A}} \def\Ab{\bar{\mathcal{A}}}
 
\def\B{\mathcal{B}} 

\def\F{\mathcal{F}}
\def\G{\mathcal{G}}

\def\H{\mathcal{H}}

\def\cP{\mathcal{P}}

\def\cR{\mathcal{R}}
\def\inn{\mathrm{in}}
\def\out{\mathrm{out}}
\def\X{{\scriptscriptstyle\mathrm{X}}}
\def\R{{\scriptscriptstyle\mathrm{R}}}
\def\V{{\scriptscriptstyle\mathrm{V}}}

\def\s{\sigma}
\def\S{\Sigma}
\def\O{\Omega}
\def\Ob{\bar{\Omega}}

\renewcommand{\d}{\mathrm{d}}
\def\p{\partial}
\newcommand{\as}{\alpha_s}
\newcommand{\asb}{\bar{\alpha}_s}

\def\uh{\hat{u}}
\def\Uh{\hat{\mathcal{U}}}
\def\qb{\bar{q}}

\def\akt{\scriptscriptstyle\mathrm{ak_t}}
\def\ktt{\scriptscriptstyle\mathrm{k_t}}
\def\CAA{\scriptscriptstyle\mathrm{C/A}}

\def\ng{\text{ng}}
\def\cl{\text{cl}}

\def\JA{\scriptscriptstyle\mathrm{JA}}
\begin{document}

\title{Hemisphere mass up to four-loops with generalised $k_t$ algorithms}

\author{
Kamel Khelifa-Kerfa \inst{1}\fnmsep\thanks{\email{kamel.khelifakerfa@relizane-univ.dz}} \and
Mohamed Benghanem  \inst{2}\fnmsep\thanks{\email{mbenghanem@iu.edu.sa} (corresponding author)}
}

\institute{
Department of Physics, Faculty of Science and Technology \\
University Ahmed Zabana of Relizane, Relizane 48000, Algeria
\and
Department of Physics, Faculty of Science, Islamic University of Madinah,\\
Madinah 42351, Saudi Arabia
}

\date{
}

\abstract{
We compute the fixed‐order distribution of the non‐global hemisphere mass observable in \(e^+e^-\) annihilation up to four loops for various sequential recombination jet algorithms. In particular, we focus on the \(k_t\) and Cambridge/Aachen algorithms. Using eikonal theory and strong‐energy ordering of the final‐state partons, we determine the complete structure of both abelian (clustering) and non‐abelian non‐global logarithms through four loops in perturbation theory. We compare the resulting resummed expressions for both jet algorithms with the standard Sudakov form factor and demonstrate that neglecting these logarithms leads to unreliable phenomenological predictions for the observable’s distribution.
\PACS{{12.38.Bx}{}\and{12.38.Cy}{}\and{13.87.-Ce}{}\and {13.85.Fb}{}
\keywords{\,QCD, Jets, Jet algorithms}}
}

\authorrunning{K. Khelifa-Kerfa et al.}
\titlerunning{Hemisphere mass up to four-loops}

\maketitle
\section{Introduction}
\label{sec:Intro}

The study of jet substructure observables in high-energy collisions provides critical insights into
the dynamics of QCD and the nature of parton evolution. Among
these observables, the hemisphere mass distribution in $e^+ e^-$ annihilation serves as a key probe
of non-global QCD effects, where correlations between emissions in distinct angular regions
generate large logarithmic corrections that are not captured by traditional Sudakov resummation.
These effects are particularly sensitive to the choice of jet algorithm, making their precise
understanding essential for both theoretical predictions and experimental analyses.

Despite being first identified more than two decades ago \cite{Dasgupta:2001sh},
{\it non-global} QCD observables continue to pose significant theoretical challenges.
Unlike {\it global} observables, which are sensitive to radiation across the entire phase space,
non-global observables depend solely on radiation within restricted regions.
This restriction inherently generates distinct energy scales in different phase-space regions,
leading to the emergence of logarithms of the ratio of these scales.
These logarithms, termed {\it non-global logarithms},
become substantial in kinematic regimes where the scale ratio is large,
and typically attain magnitudes comparable to the leading primary collinear and soft logarithms.
Their precise treatment is therefore essential for meaningful phenomenological studies of this class of observables.

The complexity in calculating non-global logarithms stems not only from the intricate structure of the radiation amplitude, even at leading orders in perturbation theory, but also from the challenging task of tracking diverse parton configurations throughout phase space to determine which contribute and which do not. For analytical calculations, an additional difficulty lies in performing the integrations, particularly owing to the multitude of integration regions and the absence of symmetry in the radiation amplitude expressions. Consequently, non-global logarithms have been calculated only up to two-loop order (where they first appear) for over a decade, in both $e^+e^-$ and hadron-hadron collisions \cite{Dasgupta:2001sh, Dasgupta:2002bw, Appleby:2002ke, Banfi:2005gj, Banfi:2010pa, Khelifa-Kerfa:2011quw, Dasgupta:2012hg, Kerfa:2012yae, Becher:2016mmh, Banfi:2021owj, Ziani:2021dxr, Bouaziz:2022tik}. Calculations beyond two loops were first undertaken in Ref.~\cite{Schwartz:2014wha}, in the large-$N_c$ limit (where $N_c$ denotes the number of colours), and in Refs.~\cite{Khelifa-Kerfa:2015mma, Delenda:2015tbo} for full colour. Further calculations have been performed for specific non-global QCD observables \cite{Benslama:2020wib, Benslama:2023gys, Khelifa-Kerfa:2024udm, Khelifa-Kerfa:2024hwx, Caron-Huot:private}. All aforementioned references have computed only the leading non-global logarithms, except for Refs.~\cite{Kerfa:2012yae, Becher:2016mmh, Banfi:2021owj}, which extend the results to next-to-leading logarithmic accuracy, albeit only at two loop order.

The resummation of non-global logarithms progressed slowly during the first decade. At leading logarithmic accuracy and in the large-\(N_c\) limit, only two approaches were available: the Monte Carlo (MC) method of Dasgupta and Salam \cite{Dasgupta:2001sh}, and the non-linear integro-differential equation of Banfi, Marchesini and Smye (BMS) \cite{Banfi:2002hw}. Both techniques are limited to a small set of observables. During the second and early third decades, significant advances were made: resummation was extended to full colour \cite{Weigert:2003mm, Hatta:2013iba, Hagiwara:2015bia, Hatta:2020wre, DeAngelis:2020rvq, Hamilton:2020rcu, vanBeekveld:2022zhl} and to next-to-leading logarithmic accuracy \cite{Banfi:2021owj, Banfi:2021xzn, Becher:2023vrh}.

Thus far, we have addressed only calculations of non-global logarithms for final-state jets defined by fixed angular constraints, such as those in cone algorithms or the anti-$k_t$ algorithm \cite{Cacciari:2008gp}. The situation is considerably more complex for other jet algorithms, such as sequential recombination algorithms including the $k_t$ \cite{Catani:1993hr, Ellis:1993tq} and Cambridge/Aachen (C/A) \cite{Dokshitzer:1997in, Wobisch:1998wt} variants. Notably, non-global logarithms are categorised into two types: Abelian, termed {\it clustering logarithms} (CLs), and non-Abelian, referred to simply as non-global logarithms (NGLs). The infinite tower of CLs originates from primary independent emissions, which also produce the leading soft and collinear Sudakov logarithms. CLs arise from mis-cancellations between real soft large-angle emissions and their virtual corrections for specific gluon configurations. Without clustering, such configurations would be absent, and consequently no mis-cancellation would occur. CLs are thus a pure consequence of the clustering effect on non-global observables. Unlike NGLs, CLs exist only for sequential clustering algorithms other than anti-$k_t$. They were first identified in Ref.~\cite{Banfi:2005gj} in response to earlier work by Appleby and Seymour \cite{Appleby:2002ke}, which demonstrated that clustering reduces the size of NGLs. Ref.~\cite{Banfi:2005gj} showed that while clustering effects reduce NGLs, they generate another tower of large logarithms: the CLs. This illustrates that the non-global nature of an observable cannot be easily circumvented by selecting alternative jet algorithms.

At fixed order, CLs were computed up to four loops early on \cite{Delenda:2006nf} owing to the relative simplicity of the corresponding emission amplitude. Similar calculations were subsequently performed for other observables \cite{Delenda:2012mm, Kelley:2012kj, Kelley:2012zs, Benslama:2023gys, Khelifa-Kerfa:2024hwx, Khelifa-Kerfa:2024udm}. Recently, CLs have been computed up to six loops for the jet mass observable \cite{Khelifa-Kerfa:2024gyv}. NGLs for the $k_t$ algorithm have been calculated up to three loops for the azimuthal decorrelation \cite{Benslama:2023gys} and up to four loops for the dijet mass \cite{Khelifa-Kerfa:2024hwx}, both in $e^+e^-$ processes. The latter work is based on Ref.~\cite{Khelifa-Kerfa:2024roc}, which presented the analytical structure of $k_t$ clustering, at leading logarithmic accuracy, to arbitrary orders in perturbation theory.

Regarding resummation, CLs and NGLs have thus far been resummed only using the Monte Carlo code of \cite{Dasgupta:2001sh}, which is limited to a small number of observables and implements only the $k_t$ algorithm. Recently, Becher and Haag successfully resummed these logarithms for all three jet algorithms using a factorisation theorem derived within Soft-Collinear Effective Theory (SCET) \cite{Becher:2023znt}. They explicitly performed the resummation for the gaps-between-jets observable at leading-logarithmic accuracy and in the large-$N_c$ limit.

In the present paper, we extend the work of Ref.~\cite{Khelifa-Kerfa:2015mma}, which computed the hemisphere mass distribution in the anti-$k_t$ algorithm, to the $k_t$ and Cambridge/Aachen (C/A) clustering algorithms. We present fixed-order calculations with full colour dependence up to four loop order in perturbation theory, demonstrating that they exhibit an exponentiation pattern. To the best of our knowledge, this is the first time such calculations have been reported in the literature, particularly for the C/A algorithm. We show that the CLs coefficients are substantially larger than those for the jet mass observable \cite{Khelifa-Kerfa:2024hwx, Khelifa-Kerfa:2024gyv}, while NGLs are generally of comparable size or smaller. This results in CLs dominating over NGLs, leading to a positive contribution to the hemisphere mass distribution. This effect manifests as an increase in the resummed distribution, particularly in the peak (shoulder) region. A similar observation was reported for the gaps-between-jets observable with large jet radius parameters in Ref.~\cite{Delenda:2006nf} (Fig.~2, third plot with $R=1$). However, this effect is absent in other non-global observables such as the jet mass \cite{Khelifa-Kerfa:2024hwx}.

This paper is organised as follows. Section~\ref{sec:Kinematics} discusses the kinematic setup, including the definition of the hemisphere mass observable, the jet algorithms considered, and the cross-section to be computed. Section~\ref{sec:FO} presents fixed-order calculations at two, three, and four loop orders. All-orders estimates are then detailed in Section~\ref{sec:Resum}, including various comparisons among the jet algorithms. Finally, conclusions are presented in Section~\ref{sec:Conclusion}.
\section{Kinematics}
\label{sec:Kinematics}

\subsection{Observable definition}
\label{sec:ObservableDef}

Consider, for simplicity and without loss of generality, the QCD process of $e^+ e^-$ annihilation to dijets. At the partonic level, this is represented as: $e^+ + e^- \to q (p_a) + \qb (p_b) + g_1(k_1) + \dots + g_n(k_n)$, where the four-momenta of the final-state partons are given by:
\begin{align}\label{eq:Def:4Momenta}
 p_a  &= \frac{Q}{2} \left(1,0,0,1\right), \notag\\
 p_b &= \frac{Q}{2} \left(1,0,0,-1\right), \notag\\
 k_i &= \omega_i \left(1, s_i \cos\phi_i, s_i \sin\phi_i, c_i\right),
\end{align}
where $c_i \equiv \cos \theta_i$ and $s_i \equiv \sin \theta_i$ with $\theta_i$ and $\phi_i$ representing the polar and azimuthal angles of the $i^{\mathrm{th}}$ gluon, $g_i$, respectively. Here $Q$ denotes the hard scale of the process and $\omega_i$ is the energy of gluon $g_i$. All partons are massless. In the eikonal (soft) approximation with strong-energy ordering, $Q \gg \omega_1 \gg \omega_2 \gg \dots \gg \omega_n$, calculations are substantially simplified while maintaining accuracy at the single-logarithmic (SL) level. At this accuracy, recoil of the hard quarks from soft-gluon emission may safely be neglected. In the threshold limit, the $q\qb$ pair is produced back-to-back along an axis that coincides, at SL accuracy, with the thrust axis. This configuration defines two coaxial hemispheres, $\H_R$ and $\H_L$, with $\theta \in [0,\pi/2]$ for $\H_R$ and $\theta \in [\pi/2, \pi]$ for $\H_L$. We measure the mass of the right hemisphere, $\H_R$, leaving the other hemisphere unmeasured.

Following the original work on NGLs \cite{Dasgupta:2001sh}, we define the invariant (normalised) squared mass of the hemisphere as:
\begin{subequations}\label{eq:Def:HemiMass}
\begin{align}
 \rho &= \frac{\left(p_a + \sum_{i \in \H_R} k_i  \right)^2}{\left(\sum_i E_i\right)^2 } \approx \sum_{i\in\H_R} \frac{2  p_a \cdot k_i}{Q^2} = \sum_{i \in\H_R} \varrho_i, \\
\varrho_i &= \frac{2 \, p_a \cdot k_i}{Q^2} = \frac{1}{2} x_i \left(1 - c_i\right),
\end{align}
where the sum runs over all (real) emissions that, after jet algorithm application, lie within $\H_R$. The energy fraction $x_i$ is defined as $x_i = 2 \omega_i/Q$. The approximation in Eq.~(\ref{eq:Def:HemiMass}a) follows from $p_a^2 = k_i^2 = 0$ (for massless partons) and $k_i \cdot k_j \propto \omega_i \omega_j \ll Q^2$ (for soft gluons).
\end{subequations}

\subsection{Jet algorithms}
\label{sec:JetAlgs}

The set of sequential recombination jet algorithms considered in this work are defined as follows \cite{Cacciari:2011ma}:
\begin{enumerate}
\item From an initial list of final-state partons, compute the distance measures for each pair $(ij)$:
\begin{subequations}\label{eq:Def:DistanceMeasure}
\begin{align}
 d_{ij} &= \min\left(E_i^{2p}, E_j^{2p}\right) \frac{1 - \cos\theta_{ij}}{1 - \cos R}, \\
d_{iB}  &= E_i^{2p},
\end{align}
\end{subequations}
where $E_i$ denotes the energy of the $i^{\mathrm{th}}$ parton, $p$ is a continuous parameter, and $\cos\theta_{ij} \equiv c_i c_j + s_i s_j \cos\phi_{ij}$ with $\phi_{ij} \equiv \phi_i - \phi_j$. The parameter $R$ represents the jet radius.

\item Identify the minimum distance among all $d_{ij}$ and $d_{iB}$. If this minimum is a $d_{ij}$, merge partons $i$ and $j$ into a pseudo-jet by summing their four-momenta (E-scheme). If the minimum is a $d_{iB}$, classify parton $i$ as an inclusive jet and remove it from the list.

\item Repeat steps 1 and 2 until no partons remain.
\end{enumerate}
Specific values $p = -1, 0, 1$ correspond to the anti-$k_t$, Cambridge/Aachen (C/A), and $k_t$ algorithms, respectively. From Eqs.~\eqref{eq:Def:DistanceMeasure}, for anti-$k_t$ ($p=-1$), harder partons always yield smaller $d_{iB} \propto E_i^{-2}$ than softer partons. Unless
\begin{align}\label{eq:Def:ClusCondition}
 1 - \cos\theta_{ij} < 1 - \cos R,
\end{align}
the minimum distance will be $d_{iB}$, forming harder partons into inclusive jets. Thus, anti-$k_t$ jets emerge as well-separated circles of radius $R$ in the $\theta$-$\phi$ plane around hard partons, with softer partons clustered only when within $R$. The strong-energy ordering condition substantially reduces the number of distance computations required, a feature common to all jet algorithms.

The $k_t$ algorithm ($p=1$) preferentially clusters softer partons first due to their smaller $E_i^2$ values. When condition \eqref{eq:Def:ClusCondition} is satisfied for soft parton pairs, they merge first. This sequential merging of soft partons before harder ones creates complex phase-space effects where gluons initially distant from hard partons may become incorporated into their jets through successive merging with other closer soft gluons. Conversely, soft gluons near hard partons may be dragged away during clustering. The $k_t$ algorithm admits more distance-ordering possibilities than anti-$k_t$, encompassing all configurations of the latter.

The Cambridge/Aachen algorithm ($p=0$) exhibits greater complexity by considering only geometric distances, independent of parton energies. Parton pairs merge solely when satisfying \eqref{eq:Def:ClusCondition}. With the largest set of possible orderings, C/A includes all configurations present in both anti-$k_t$ and $k_t$.

These algorithmic mechanisms apply generally. For the hemisphere geometry considered here ($R = \pi/2$), soft emissions reside exclusively in either the right or left hemisphere, precluding additional jets. Thus, $k_t$ and C/A effects reduce to gluon migration across the inter-hemisphere boundary without separate jet formation. Eq.~\eqref{eq:Def:DistanceMeasure} simplifies to:
\begin{align}
 d_{ij} = \min \left(E_i^{2p}, E_j^{2p}\right) \left(1 - \cos\theta_{ij}\right),
\end{align}
with $d_{iB}$ irrelevant since only the hard $q\qb$ pair can form jets (which are the right and left hemispheres).

Recalling from Eq.~\eqref{eq:Def:HemiMass} that the hemisphere mass depends critically on gluon assignments to $\H_R$ after jet clustering, we anticipate distinct observable distributions for different algorithms. This will be explored in detail in Sec. \ref{sec:FO}.

\subsection{Observable distribution}
\label{sec:ObservableDist}

We are interested in computing the integrated cross‐section for the non‐global hemisphere mass distribution:
\begin{align}\label{eq:Def:Sigma-Integrated}
 \S(\rho) = \int \frac{1}{\s_0} \frac{\d \s}{\d\varrho} \,\Theta\bigl[\rho - \varrho(k_1,\dots,k_n)\bigr]\,\Xi(k_1,\dots,k_n),
\end{align}
where \(\s_0\) is the Born cross‐section and \(\d\s/\d\varrho\) is the differential cross‐section in the normalised hemisphere mass \(\varrho(k_1,\dots,k_n)\), defined in Eq.~\eqref{eq:Def:HemiMass}. The integration is over all parton configurations that contribute to the hemisphere mass by an amount less than the {\it veto} \(\rho\). The {\it clustering function}, \(\Xi(k_1,\dots,k_n)\), encodes the phase‐space constraints arising from the application of a given jet algorithm to the set of final‐state partons contributing to the hemisphere mass at a given loop order in perturbation theory. The perturbative fixed‐order expansion of \(\S(\rho)\) can be written as:
\begin{align}\label{eq:Def:Sigma-PTexpand}
 \S(\rho) = 1 + \S_1(\rho) + \S_2(\rho) + \cdots,
\end{align}
where the \(m\)$^\text{th}$ order contribution can be cast in the following form \cite{Khelifa-Kerfa:2015mma, Khelifa-Kerfa:2024hwx}:
\begin{multline}\label{eq:Def:Sigma-m}
\S_m(\rho) = \int_{x_1 > \cdots >x_m} \prod_{i=1}^m \d\Phi_i \sum_X \Uh_m\,\W_{1\ldots m}^{\X}\,\Xi_m(k_1,\dots,k_n),
\end{multline}
with strongly ordered energy fractions \(x_1>\cdots>x_m\) and the phase‐space element
\begin{align}
 \d\Phi_i = \asb \,\frac{\d x_i}{x_i}\,\d c_i\,\frac{\d\phi_i}{2\pi},
\end{align}
where \(\asb = \as/\pi\). The eikonal amplitudes squared, \(\W_{1\ldots m}^\X\), for the emission of \(m\) soft, energy-ordered gluons off the primary \(q\bar q\) partons in configuration \(X\) were derived in Ref.~\cite{Delenda:2015tbo}. The symbol \(X\) represents a real (R) or virtual (V) assignment of each emitted gluon. For instance, at one loop the emitted gluon \(k_1\) can be real or virtual, \(X\in\{R,V\}\), yielding \(\W_1^R\) or \(\W_1^V\). At two loops \(X\in\{RR, RV, VR, VV\}\), with corresponding amplitudes give by \(\W_{12}^{\R\R}, \W_{12}^{\V\R}, \W_{12}^{\R\V}, \W_{12}^{\V\V}\). For further details on these amplitudes and their properties, see Ref.~\cite{Delenda:2015tbo}. The sum in \eqref{eq:Def:Sigma-m} runs over all real/virtual configurations at a given loop order.

The term \(\Uh_m\), known as the {\it measurement operator} \cite{Schwartz:2014wha, Khelifa-Kerfa:2015mma}, selects gluon emissions whose contributions to the hemisphere mass are less than the veto \(\rho\). In the strong‐energy–ordering limit, it factorises into
\begin{subequations}
\begin{align}
\Uh_m &= \prod_{i=1}^m \uh_i, \\
\uh_i &= \Theta_i^\V + \Theta_i^\R\bigl[\Theta_i^\out + \Theta_i^\inn\,\Theta(\rho - \varrho_i)\bigr]
       = 1 - \Theta_i^\rho\,\Theta_i^\inn\,\Theta_i^\R,
\label{eq:Def:MeasOperator}
\end{align}
\end{subequations}
where \(\Theta_i^{\R(\V)}=1\) if gluon \(k_i\) is real (virtual), and \(\Theta_i^{\out(\inn)}=1\) if \(k_i\) is outside (inside) the measured hemisphere \(\H_R\). Also \(\Theta_i^\rho \equiv \Theta(\varrho_i - \rho)\). The expression for \(\uh_i\) shows that an emission must be real, inside \(\H_R\), and have \(\varrho_i<\rho\) to contribute to \(\S_m(\rho)\). Note the complementarity relations
\begin{align}
 \Theta_i^\inn + \Theta_i^\out = 1, \qquad \Theta_i^\R + \Theta_i^\V = 1.
\end{align}

In the next section we present fixed‐order calculations at one, two, three, and four loops for both the \(k_t\) and C/A jet algorithms. The corresponding calculations for the anti‐\(k_t\) algorithm were performed a decade ago \cite{Khelifa-Kerfa:2015mma}; we shall, nonetheless, report them here for comparison.

\section{Fixed-order calculations}
\label{sec:FO}

We begin by presenting the one loop calculation. Recall from the definition of the generalised $k_t$ algorithms, Sec. \ref{sec:JetAlgs}, that all of them work identically at this order. The sum over possible gluon configurations in Eq. \eqref{eq:Def:Sigma-m}, for $m=1$, reads:
\begin{align}\label{eq:1loop:uWX}
 \sum_\X \uh_1 \W_1^\X = \uh_1 \W_1^\R + \uh_1 \W_1^\V = -\Theta_i^\rho\,\Theta_i^\inn\, \W_1^\R,
\end{align}
where we have used the formula of the measurement operator \eqref{eq:Def:MeasOperator} to arrive at the last equality.
Substituting back into Eq. \eqref{eq:Def:Sigma-m} we obtain, for the one loop integrated hemisphere mass distribution,
\begin{align}\label{eq:1loop:Sigma}
\S_1(\rho) &= -\int \d\Phi_1\, \Theta_1^\rho\, \Theta_1^\inn\, \W_1^\R, \notag\\
		   &= -\asb\,\CF \int_0^1 \frac{\d x_1}{x_1} \int_0^1 \d c_1 \int_0^{2\pi} \frac{\d\phi_1}{2\pi}\times \notag \\
		   &\times  w_{ab}^1\, \Theta\left[x_1(1-c_1) - 2 \rho\right],
\end{align}
where $\CF = (N_c^2-1)/2N_c$ is the Casimir colour factor in the fundamental representation of SU$(N_c)$. The one loop eikonal amplitudes squared are $\W_k^\R = \CF\, w_{ij}^k$ and $\W_k^\V = -\W_k^\R$, with the one loop antenna function, $w_{ij}^k$, given by:
\begin{align}\label{eqL1loop:Antenna}
 w_{ij}^k = \omega_k^2 \frac{(p_i \cdot p_j)}{(p_i \cdot k)(p_j \cdot k)} = \omega_k^2 \frac{(ij)}{(ik)(kj)}.
\end{align}
Substituting for the four momenta in Eq.~\eqref{eq:Def:4Momenta}, we obtain $w_{ab}^1 = 2/(1-c_1^2)$. To include contributions from hard collinear emissions, we simply replace $1/x_1$ by $\bigl(1+(1-x_1)^2\bigr)/(2x_1)$ in Eq.~\eqref{eq:1loop:Sigma}. Performing the integration, we find at SL accuracy
\begin{align}\label{eq:1loop:Sigma-Final}
 \S_1(\rho) = -\CF\,\asb \left[\tfrac12\,L^2 - \tfrac34\,L\right],
\end{align}
where $L = \ln(1/\rho)$. As is well known, the double logarithmic contribution arises from regions of phase space where emissions are both soft and collinear to the emitting hard partons. It is also known that the resummation of such large logarithms to all orders is given by the exponential of the one loop result (the Sudakov form factor) \cite{Catani:1992ua}. Details about the latter will be presented in Sec.~\ref{sec:Resum}.

The difference between the various jet algorithms starts to appear at two loop order, which we shall present in the following section.

\subsection{Two loops}
\label{sec:2loop}

For the emission of two soft energy‐ordered gluons, \(k_1\) and \(k_2\), the sum over all possible gluon configurations for the product of the projection operator and eikonal amplitudes squared in Eq.~\eqref{eq:Def:Sigma-m} (\(m=2\)) is \cite{Khelifa-Kerfa:2024roc}:
\begin{align}\label{eq:2loop:uWX}
 \sum_\X \Uh_2 \W_{12}^\X = -\Theta_1^\rho \,\Theta_2^\rho\, \Theta_2^\inn
 \Bigl[\W_{12}^{\V\R} + \Theta_1^\out\,\W_{12}^{\R\R}\,\Xi_2(k_1,k_2)\Bigr],
\end{align}
where the clustering functions \(\Xi_2(k_1,k_2)\) for the three jet algorithms are:
\begin{align}\label{eq:2loop:ClusFun}
 \Xi_2^{\akt} = 1,
 \qquad
 \Xi_2^{\ktt} = \Xi_2^{\CAA} = \Ob_{12},
\end{align}
with the constraint \(\Ob_{ij} = \Theta(d_{ij} - d_{ja}) = 1 - \O_{ij}\), which forces the distance between partons \(i,j\) to exceed that between parton \(j\) and the hard quark \(a\). Hence, in our case, gluons \(k_1\) and \(k_2\) are not clustered together. Note that at this order the \(k_t\) and C/A algorithms share the same clustering function.

Using \(\Theta_1^\inn + \Theta_1^\out = 1\) and \(\O_{12} + \Ob_{12} = 1\), together with the explicit eikonal amplitudes squared \cite{Delenda:2015tbo}:
\begin{align}\label{eq:2loop:EikAmps}
 &\W_{12}^{\R\R} = \W_1^\R \W_2^\R + \oW_{12}^{\R\R},
 \quad \W_{12}^{\R\V} = -\W_{12}^{\R\R}, \notag\\
 &\W_{12}^{\V\R} = -\W_1^\R \W_2^\R,
 \qquad \W_{12}^{\V\V} = -\W_{12}^{\V\R},
\end{align}
we can expand Eq.~\eqref{eq:2loop:uWX} to:
\begin{multline}\label{eq:2loop:uWX-B}
 \sum_\X \Uh_2 \W_{12}^\X = -\Theta_1^\rho\,\Theta_2^\rho \Bigl[
    -\Theta_1^\inn\,\Theta_2^\inn\,\W_1^\R \W_2^\R
    - \Theta_1^\out\,\Theta_2^\inn\times \\[-4pt]
    \times \W_1^\R \W_2^\R\,\O_{12} + \Theta_1^\out\,\Theta_2^\inn\,\oW_{12}^{\R\R}\,\Ob_{12}
 \Bigr],
\end{multline}
where for the anti-\(k_t\) algorithm \(\O_{12}=0\) and \(\Ob_{12}=1\). The first term on the RHS corresponds to successive independent emissions of soft gluons, all inside the measured region; it is always present, for all jet algorithms and at any loop order, and represents the expansion of the Sudakov form factor at this order. Substituting it back into Eq.~\eqref{eq:Def:Sigma-Integrated} yields
\begin{multline}
 \frac{1}{2!}\Bigl(-\int \d\Phi_1\,\Theta_1^\rho\,\Theta_1^\inn\,\W_1^\R\Bigr)
 \times \Bigl(-\int \d\Phi_2\,\Theta_2^\rho\,\Theta_2^\inn\,\W_2^\R\Bigr) \\[-4pt]
 = \frac{1}{2!}\,(\S_1)^2,
\end{multline}
where we have relaxed the strong ordering \(x_1>x_2\) and included the factor \(1/2!\) to avoid double counting (the integrand is symmetric under \(1\leftrightarrow2\)).

The second term on the RHS of \eqref{eq:2loop:uWX-B} is absent for the anti-$k_t$ algorithm and present for the $k_t$ and C-A algorithms. It represents the contribution from clustering logarithms. Substituting it back into Eq.~\eqref{eq:Def:Sigma-Integrated}, we write
\begin{align}\label{eq:2loop:CLs}
 \S_{2,\cl}^{\ktt}(\rho) =  \S_{2,\cl}^{\CAA}(\rho) = \int_{1_\out} \int_{2_\inn} \O_{12}\, \W_1^\R \W_2^\R = \asb^2 \frac{L^2}{2!} \CFsq \,\F_{2},
\end{align}
where we have introduced the notation
\begin{align}
 \int_{i_{\out(\inn)}} = \int \d\Phi_i \, \Theta_i^\rho \, \Theta_i^{\out(\inn)},
\end{align}
and the two loop clustering coefficient, \(\F_2\):
\begin{align}\label{eq:2loop:F2}
\F_2 &= \int_{1_\out} \int_{2_\inn} \O_{12}\, w_{ab}^1\,w_{ab}^2, \notag\\
	 &= \int_{-1}^0 \d c_1 \int_0^1 \d c_2 \iint_0^{2\pi} \frac{\d\phi_1}{2\pi} \frac{\d\phi_2}{2\pi} \, \frac{4\, \Theta\left(\cos\theta_{12} - c_2 \right)}{(1-c_1^2)(1-c_2^2)},
\notag\\
&= 1.2337 \approx \frac{\pi^2}{8},
\end{align}
where the last approximation has been estimated using the \texttt{PolyLogTools} package \cite{Duhr:2019tlz}. We note that the above value of \(\F_2\) is much larger than that reported for the jet mass observable in the limit \(R\to 0\) \cite{Banfi:2010pa, Khelifa-Kerfa:2011quw, Delenda:2012mm, Khelifa-Kerfa:2024hwx}, namely \(\lim_{R\to 0}\F_2(R)=0.183\). This difference may be due to the larger phase space available for primary independent emissions in the hemisphere observable compared to the jet mass, which is constrained by the jet radius \(R\).

The third term on the RHS of Eq.~\eqref{eq:2loop:uWX-B} corresponds to the NGLs contribution at this order:
\begin{subequations}\label{eq:2loop:NGLs}
\begin{align}
 \S_{2,\ng}^{\akt}(\rho) &= -\int_{1_\out} \int_{2_\inn} \oW_{12}^{\R\R}, \\
 \S_{2,\ng}^{\ktt}(\rho) = \S_{2,\ng}^{\CAA}(\rho) &= -\int_{1_\out} \int_{2_\inn} \oW_{12}^{\R\R}\, \Ob_{12},
\end{align}
\end{subequations}
with the irreducible eikonal amplitude squared given, in terms of the two loop antenna \(\A_{ab}^{ij}\), by \cite{Delenda:2012mm}:
\begin{align}\label{eq:2loop:oW}
 \oW_{ij}^{\R\R} = \frac{1}{2} \CF\CA\, \A_{ab}^{ij}, \qquad
 \A_{ab}^{ij} = w_{ab}^i \bigl(w_{ai}^j + w_{bi}^j - w_{ab}^j \bigr).
\end{align}
At SL accuracy, we can write these contributions, for a given jet algorithm (JA), as:
\begin{align}\label{eq:2loop:S2}
 \S_{2,\ng}^{\JA}(\rho) = -\asb^2 \,\frac{L^2}{2!}\, \CF\CA\, \G_2^{\JA}.
\end{align}
The two loop NGLs coefficient, \(\G_2\), has been computed for the anti-$k_t$ algorithm in our previous work \cite{Khelifa-Kerfa:2015mma}: \(\G_2^{\akt} = \zeta_2 = 1.645\). The \(k_t\) and C/A results for the hemisphere mass observable, which have not been computed before, are found to be \(\G_2^{\ktt} = \G_2^{\CAA} = 0.2056 \approx \zeta_2/8\), representing a reduction of more than 85\% compared to the anti-$k_t$ case. Combining the CLs and NGLs contributions for \(k_t\) and C/A at this order, one obtains \(\bigl(\CFsq \F_2 - \CF\CA \G_2\bigr)/2! = 0.685\). For the anti-$k_t$ algorithm the NGLs term is given by \(-\CF\CA \G_2^{\akt}/2! = -2\zeta_2 = -3.29\), indicating that clustering reduces NGLs by approximately 80\%, in agreement with earlier observations \cite{Appleby:2002ke, Banfi:2005gj, Banfi:2010pa, Khelifa-Kerfa:2011quw, Khelifa-Kerfa:2024roc}.

Moreover, as for the clustering coefficient \(\F_2\), the NGLs coefficient \(\G_2\) for \(k_t\) clustering differs from that reported for the jet mass observable in the \(R\to 0\) limit \cite{Khelifa-Kerfa:2011quw, Khelifa-Kerfa:2024hwx}. By contrast, in the case of the anti-$k_t$ algorithm (no clustering) the hemisphere mass result coincides with the \(R\to0\) limit of the jet mass observable. This discrepancy arises because clustering algorithms depend on the geometric phase‐space distribution of final‐state partons. Since NGLs and CLs are primarily an {\it edge} (or {\it boundary}) effect \cite{Dasgupta:2001sh, Dasgupta:2002bw, Khelifa-Kerfa:2011quw}, any boundary between the in and out regions yields NGLs. Without clustering, that boundary is identical for jet mass and hemisphere observables; clustering, however, {\it distorts} the boundary by dragging gluons in or out, and this distortion differs between jet and hemisphere observables, leading to distinct CLs and NGLs coefficients.

Finally, we observe that the hemisphere CLs coefficient exceeds that for the jet mass by more than a factor of six, while the NGLs coefficient is over three times smaller. Hence, at two loops the combined effect on the hemisphere mass distribution is positive, in contrast to the jet mass case. The full distribution may be expressed, up to SL accuracy, as:
\begin{align}\label{eq:2loop:Sigma-FinalForm}
 \S_2^{\JA}(\rho) = \frac{1}{2!} (\S_1)^2 + \S_{2,\cl}^{\JA} + \S_{2,\ng}^{\JA}.
\end{align}

\subsection{Three loops}
\label{sec:3loop}

For the emission of three soft energy‐ordered gluons, \(k_1\), \(k_2\) and \(k_3\), the sum over all possible gluon configurations for the eikonal amplitudes squared in \eqref{eq:Def:Sigma-Integrated} (\(m=3\)) is given by \cite{Khelifa-Kerfa:2024roc}:
\begin{align}\label{eq:3loop:uWX}
\sum_{\X} \Uh_3 \,\W_{123}^{\X} &= -\prod_{i=1}^3 \Theta_i^\rho\,\Theta_3^\inn \Bigl[
 \W_{123}^{\V\V\R} + \Theta_1^\out\,\W_{123}^{\R\V\R}\,\Ob_{13} \notag\\
&\;\;+\;\Theta_2^\out\,\W_{123}^{\V\R\R}\,\Ob_{23} \notag\\
&\;\;+\;\Theta_1^\out\bigl(\Theta_2^\out + \Theta_2^\inn\,\O_{12}\bigr)\,
 \W_{123}^{\R\R\R}\,\Ob_{13}\,\Ob_{23}
\Bigr] \notag\\ &\;\;+ \cP_3^{\CAA},
\end{align}
where, for the anti-\(k_t\) algorithm, \(\O_{ij}=0\) and \(\Ob_{ij}=1\), and the amplitude squared \(\cP_3^{\CAA}\) is nonzero only for the C/A algorithm. Its full derivation will be presented elsewhere \cite{CAclus}; it reads
\begin{multline}\label{eq:3loop:P-CA}
 \cP_3^{\CAA} = -\prod_{i=1}^3 \Theta_i^\rho\,\Theta_3^\inn\,\Theta_1^\out\,\O_{12}\,\Ob_{13}\,\O_{23}\,\Delta_{123}\\
 \times \bigl[\W_{123}^{\R\R\R} + \W_{123}^{\R\V\R} + \W_{123}^{\V\V\R}\bigr],
\end{multline}
with \(\Delta_{ijk} \equiv \Theta(d_{jk} - d_{ij})\).

The eikonal amplitudes squared are \cite{Delenda:2012mm}:
\begin{subequations}\label{eq:3loop:EikAmps}
\begin{align}
 \W_{123}^{\R\R\R} &= \prod_{i=1}^3 \W_i^\R + \prod_{i=1}^3 \W_i^\R\,\oW_{jk}^{\R\R} + \oW_{123}^{\R\R\R}, \\
 \W_{123}^{\R\V\R} &= -\prod_{i=1}^3 \W_i^\R - \prod_{j=2}^3 \W_j^\R\,\oW_{1k}^{\R\R} + \oW_{123}^{\R\V\R}, \\
 \W_{123}^{\V\R\R} &= -\prod_{i=1}^3 \W_i^\R - \W_1^\R\,\oW_{23}^{\R\R}, \\
 \W_{123}^{\V\V\R} &= \prod_{i=1}^3 \W_i^\R,
 \quad
 \W_{123}^{\X_1\X_2\V} = -\W_{123}^{\X_1\X_2\R}.
\end{align}
\end{subequations}
The one and two loop amplitudes squared have already been given above.
The three‐loop irreducible amplitudes are \cite{Delenda:2015tbo}:
\begin{subequations}\label{eq:3loop:EikAmps-Irred}
\begin{align}
 \oW_{123}^{\R\R\R} &= \tfrac{1}{4}\,\CF\,\CAsq
   \bigl[\A_{ab}^{12}\,\Ab_{ab}^{13} + \B_{ab}^{123}\bigr], \\
 \oW_{123}^{\R\V\R} &= -\tfrac{1}{4}\,\CF\,\CAsq\,\A_{ab}^{12}\,\Ab_{ab}^{13},
\end{align}
\end{subequations}
where \(\Ab_{ab}^{ij}=\A_{ab}^{ij}/w_{ab}^i\) and the three loop antenna, $\B_{ab}^{ijk}$, is defined as:
\begin{align}
\B_{ab}^{ijk}=w_{ab}^i\bigl[\A_{ai}^{jk}+\A_{bi}^{jk}-\A_{ab}^{jk}\bigr].
\end{align}
Substituting into \eqref{eq:3loop:uWX} and using \(\Theta_i^\out+\Theta_i^\inn=1\) and \(\O_{ij}+\Ob_{ij}=1\), we obtain:
\begin{align}\label{eq:3loop:uWX-B}
 \sum_{\X} \Uh_3 \W_{123}^{\X} &= -\prod_{i=1}^3 \Theta_i^\rho\,\Theta_3^\inn \Bigl[
   \Theta_1^\inn\,\Theta_2^\inn\,\W_{123}^{\V\V\R} \notag\\
&\;\;+\,\Theta_1^\out\,\Theta_2^\out\bigl(\Ob_{13}\,\Ob_{23}\,\W_{123}^{\R\R\R}
        +\Ob_{23}\,\W_{123}^{\V\R\R} \notag\\
&\quad\quad\;+\;\Ob_{13}\,\W_{123}^{\R\V\R} + \W_{123}^{\V\V\R}\bigr) \notag\\
&\;\;+\,\Theta_1^\out\,\Theta_2^\inn\bigl(\O_{12}\,\Ob_{13}\,\Ob_{23}\,\W_{123}^{\R\R\R}
        +\Ob_{13}\,\W_{123}^{\R\V\R} \notag\\
&\quad\quad\;+\;\W_{123}^{\V\V\R}\bigr) \notag\\
&\;\;+\,\Theta_1^\inn\,\Theta_2^\out\bigl(\Ob_{23}\,\W_{123}^{\V\R\R}
        + \W_{123}^{\V\V\R}\bigr)
\Bigr]
+ \cP_3^{\CAA}.
\end{align}
The first term, where all gluons lie inside the measured hemisphere, represents the primary independent emission part and is contained in the Sudakov form factor. The remaining terms include interference and new contributions. For example, the term proportional to \(\Theta_1^\inn\,\Theta_2^\out\) simplifies to
\begin{align}
-\prod_{i=1}^3 \Theta_i^\rho\,\Theta_3^\inn\,\Theta_1^\inn\,\Theta_2^\out
\bigl[\O_{23}\,\W_{123}^{\V\V\R} - \Ob_{13}\,\W_1^\R\,\oW_{23}^{\R\R}\bigr].
\end{align}
The first part integrates to \(L^2/2!\times\F_2\,L^2\), representing interference between the one loop cross-section \(\S_1\) and the CLs part of the two loop cross-section \(\S_{2,\cl}\). The second part integrates to \(L^2/2!\times\G_2\,L^2\), representing interference between \(\S_1\) and the NGLs part of the two loop cross-section \(\S_{2,\ng}\). Upon integration of all terms one finds that the three loop distribution assumes the form:
\begin{align}\label{eq:3loop:Sigma-FinalForm}
 \S_3^{\JA}(\rho) = \frac{1}{3!} (\S_1)^3 + \S_1 \bigl[\S_{2,\cl}^{\JA} + \S_{2,\ng}^{\JA}\bigr]
 + \S_{3,\cl}^{\JA} + \S_{3,\ng}^{\JA}.
\end{align}

The three loop CLs contribution may be written as:
\begin{subequations}
\begin{align}\label{eq:3loop:CLs}
 \S_{3,\cl}^{\JA}(\rho) &= -\asb^3\,\frac{L^3}{3!}\,\CFcub\,\F_3^{\JA},
\end{align}
where the three loop CLs coefficient is given, for the $k_t$ clustering, by the following integral \cite{Khelifa-Kerfa:2024hwx, Khelifa-Kerfa:2024hwz}:
\begin{multline}\label{eq:3loop:F3-KT}
 \F_3^{\ktt} = \Biggl[
   \int_{1_\out}\!\int_{2_\out}\!\int_{3_\inn} \O_{13}\,\O_{23} \\
 \quad + \int_{1_\out}\!\int_{2_\inn}\!\int_{3_\inn} \O_{12}\bigl(-1 + \Ob_{13}\,\Ob_{23}\bigr)
 \Biggr]\,w_{ab}^1\,w_{ab}^2\,w_{ab}^3 \\
 = 1.288.
\end{multline}
\end{subequations}
Although $\F_3^{\ktt}$ is larger than the two loop CLs coefficient $\F_2^{\ktt}$, it is nonetheless suppressed by a factor of $\CF/3 = 4/9$; that is, $\CFcub\,\F_3^{\ktt}/3! = 0.51$, which is less than $\CFsq\,\F_2^{\ktt}/2! = 1.1$, hence the convergence of the perturbative series is preserved.

For the C/A algorithm, the three loop CLs coefficient may be written as:
\begin{align}\label{eq:3loop:F3-CA}
 \F_3^{\CAA} &= \F_3^{\ktt} + f_3^{\CAA},
\end{align}
where
\begin{align}
 f_3^{\CAA} &= \int_{1_\out}\!\int_2\!\int_{3_\inn}\,\O_{12}\,\Ob_{13}\,\O_{23}\,\Delta_{123}\,w_{ab}^1\,w_{ab}^2\,w_{ab}^3 = 0.075.
\end{align}
We note that the constraints on $k_3$ restrict $k_2$ to the region inside the measured right hemisphere, i.e.\ $1>c_2>0$. Thus the three loop C/A CLs coefficient reads $\F_3^{\CAA}=1.363$. Comparing to the two loop results, we see that $\CFcub\,\F_3^{\CAA}/3! = 0.54$, which is again smaller than the two loop coefficient and thus the perturbative series remains convergent.

The three loop NGLs contribution, \(\S_{3,\ng}^{\JA}\), to the integrated hemisphere mass cross‐section may be written in the following form \cite{Khelifa-Kerfa:2024hwx}:
\begin{align}\label{eq:3loop:NGLs}
 \S_{3,\ng}^{\JA}(\rho) &= +\asb^3\,\frac{L^3}{3!}\,\bigl[\CFsq\,\CA\,\G_{3,a}^{\JA} + \CF\,\CAsq\,\G_{3,b}^{\JA}\bigr],
\end{align}
where it should be noted that clustering affects all colour channels. For the anti-\(k_t\) algorithm, the first term vanishes and the second is \(\G_{3,b}^{\akt}=\zeta_3\) \cite{Khelifa-Kerfa:2015mma, Khelifa-Kerfa:2024hwx}. For the \(k_t\) clustering they are given by \cite{Khelifa-Kerfa:2024hwx}:
\begin{subequations}
\begin{align}\label{eq:3loop:G3a-KT}
 \G_{3,a}^{\ktt} &= \tfrac{1}{2}\Biggl[
   \int_{1_\out}\!\int_{2_\out}\!\int_{3_\inn}\Bigl(\O_{13}\,\Ob_{23}\,w_{ab}^1\,\A_{ab}^{23} + \notag\\
    &\qquad + \Ob_{13}\,\O_{23}\,[w_{ab}^2\,\A_{ab}^{13}+w_{ab}^3\,\A_{ab}^{12}]\Bigr)\notag\\
 &\quad - \int_{1_\out}\!\int_{2_\inn}\!\int_{3_\inn}\Bigl(\O_{12}\,\Ob_{13}\,\Ob_{23}\,(w_{ab}^1\,\A_{ab}^{23}+w_{ab}^2\,\A_{ab}^{13})+\notag\\
 &\quad\quad + (\O_{13} + \O_{12}(-1+\Ob_{13}\,\Ob_{23}))\,w_{ab}^3\,\A_{ab}^{12}\Bigr)
 \Biggr] = 0.610,
\end{align}
which approximates to about \(\zeta_3/2\), and
\begin{align}\label{eq:3loop:G3b-KT}
 \G_{3,b}^{\ktt} &= \frac{1}{4}\Biggl[
   \int_{1_\out}\!\int_{2_\out}\!\int_{3_\inn}\Bigl(\Ob_{13}\,\O_{23}\,\A_{ab}^{12}\,\Ab_{ab}^{13}
     - \Ob_{13}\,\Ob_{23}\,\B_{ab}^{123}\Bigr)\notag\\
 &\quad + \int_{1_\out}\!\int_{2_\inn}\!\int_{3_\inn}\Bigl(\Ob_{13}(1-\O_{12}\,\Ob_{23})\,\A_{ab}^{12}\,\Ab_{ab}^{13}-
     \notag\\ &\quad- \O_{12}\,\Ob_{13}\,\Ob_{23}\,\B_{ab}^{123}\Bigr)
 \Biggr] = 0.032,
\end{align}
\end{subequations}
which approximates to about \(\zeta_3/38\). For the C/A algorithm, the two parts are given by:
\begin{subequations}
\begin{align}
 \G_{3,a}^{\CAA} &= \G_{3,a}^{\ktt} + g^{\CAA}_{3,a}, \\
 \G_{3,b}^{\CAA} &= \G_{3,b}^{\ktt} + g^{\CAA}_{3,b},
\end{align}
\end{subequations}
with the additional C/A‐clustering coefficients \cite{CAclus}:
\begin{align}
 g^{\CAA}_{3,a} &= \frac{1}{2}\int_{1_\out}\!\int_2\!\int_{3_\inn}\O_{12}\,\Ob_{13}\,\O_{23}\,\Delta_{123}\,w_{ab}^1\,\A_{ab}^{23} = 0.045,\\
 g^{\CAA}_{3,b} &= \frac{1}{4}\int_{1_\out}\!\int_2\!\int_{3_\inn}\O_{12}\,\Ob_{13}\,\O_{23}\,\Delta_{123}\,\B_{ab}^{123}    = 0.166.
\end{align}
Therefore \(\G_{3,a}^{\CAA}=0.655\) and \(\G_{3,b}^{\CAA}=0.198\). Adding CLs and NGLs at this order gives
\begin{align}
 &\tfrac{1}{3!}\,\CF\,\CAsq\,\G_3^{\akt} = 2.41,\notag\\
 &\tfrac{1}{3!}\bigl(-\CFcub\,\F_3^{\ktt} + \CFsq\,\CA\,\G_{3,a}^{\ktt} + \CF\,\CAsq\,\G_{3,b}^{\ktt}\bigr) = 0.10,\notag\\
 &\tfrac{1}{3!}\bigl(-\CFcub\,\F_3^{\CAA} + \CFsq\,\CA\,\G_{3,a}^{\CAA} + \CF\,\CAsq\,\G_{3,b}^{\CAA}\bigr) = 0.44.
\end{align}
It is evident that clustering induces a substantial reduction in the non‐global contribution to the hemisphere mass distribution, more severe for \(k_t\) clustering. C/A contributions exceed those of \(k_t\), likely due to the larger phase space from different orderings of the measure distances \(d_{ij}\) (see Sec.~\ref{sec:JetAlgs}). Furthermore, whereas the anti-\(k_t\) NGLs coefficient matches the \(R\to0\) limit of the jet mass observable, the \(k_t\) and C/A results for both CLs and NGLs differ between the two observables—a behaviour already noted at two loops and explained therein.

\subsection{Four loops}
\label{sec:4loop}

For the emission of four soft, energy-ordered gluons, the sum over the possible configurations of the various eikonal amplitudes squared in \eqref{eq:Def:Sigma-m} ($m=4$) is given by \cite{Khelifa-Kerfa:2024roc}:
\begin{align}\label{eq:4loop:uWX}
\sum_{\X} \Uh_4 \W_{1234}^{\X} &= -\prod_{i=1}^{4} \Theta_i^\rho\,\Theta_4^\inn \Big[
\W_{1234}^{\V\V\V\R} + \Theta_1^\out\,\Ob_{14}\,\W_{1234}^{\R\V\V\R} + \notag\\
&+  \Theta_2^\out\,\Ob_{24}\,\W_{1234}^{\V\R\V\R} +  \Theta_3^\out\,\Ob_{34}\,\W_{1234}^{\V\V\R\R} +
\notag\\
&+ \Theta_1^\out\left(\Theta_2^\out + \Theta_2^\inn \O_{12} \right) \Ob_{14} \Ob_{24}  \W_{1234}^{\R\R\V\R} + \notag\\
&+ \Theta_1^\out\left(\Theta_3^\out + \Theta_3^\inn \O_{13} \right) \Ob_{14} \Ob_{34}  \W_{1234}^{\R\V\R\R} + \notag\\
&+ \Theta_2^\out\left(\Theta_3^\out + \Theta_3^\inn \O_{23} \right) \Ob_{24} \Ob_{34}  \W_{1234}^{\V\R\R\R} + \notag\\
&+ \Theta_1^\out \left(\Theta_2^\out + \Theta_2^\inn \O_{12} \right) \times \notag\\
& \times \left(\Theta_3^\out + \Theta_3^\inn \left[\O_{23} + \O_{13} \Ob_{23} \right] \right) \times \notag\\
& \times \Ob_{14} \Ob_{24} \Ob_{34}  \W_{1234}^{\R\R\R\R}
\Big] + \cP_4^{\CAA}.
\end{align}
Here, as usual, for the anti-$k_t$ algorithm all clustering constraints vanish ($\O_{ij} = 0$). The expressions for the eikonal amplitudes squared are given in Ref.~\cite{Delenda:2015tbo}, and the C/A clustering–specific contribution, $\cP_4^{\CAA}$, is detailed in Ref.~\cite{CAclus}. By following the procedure outlined at two and three loop orders—particularly using $\O_{ij} + \Ob_{ij} = 1$—the above formula can be decomposed into eight phase‐space regions, only four of which yield new contributions. These share the factor $\Theta_1^\out\,\Theta_4^\inn$ together with the four possible in/out configurations of gluons $k_2$ and $k_3$:
\[
\Theta_2^\out\,\Theta_3^\out,\quad
\Theta_2^\out\,\Theta_3^\inn,\quad
\Theta_2^\inn\,\Theta_3^\out,\quad
\Theta_2^\inn\,\Theta_3^\inn.
\]
The remaining four regions, in which the hardest gluon $k_1$ is always inside the hemisphere, represent {\it interference} contributions.

The hemisphere mass distribution at four loops may be cast in a form analogous to that at two and three loops, Eqs.~\eqref{eq:2loop:Sigma-FinalForm} and \eqref{eq:3loop:Sigma-FinalForm}:
\begin{align}\label{eq:4loop:Sigma-FinalForm}
 \S_4^{\JA}(\rho) &= \frac{1}{4!} \bigl(\S_1\bigr)^4
   + \S_1\,\bigl[\S_{3,\cl}^{\JA} + \S_{3,\ng}^{\JA}\bigr] \notag\\
&\quad + \frac{1}{2!} \bigl(\S_1\bigr)^2\,\bigl[\S_{2,\cl}^{\JA} + \S_{2,\ng}^{\JA}\bigr]
   + \frac{1}{2!} \bigl(\S_{2,\cl}^{\JA}\bigr)^2 \notag\\
&\quad + \frac{1}{2!} \bigl(\S_{2,\ng}^{\JA}\bigr)^2
   + \S_{4,\cl}^{\JA} + \S_{4,\ng}^{\JA},
\end{align}
where only \(\S_{4,\cl}^{\JA}\) and \(\S_{4,\ng}^{\JA}\) are the new (CLs and NGLs) contributions at this order.

The CLs contribution, \(\S_{4,\cl}^{\JA}\), to the hemisphere mass for a given jet algorithm (JA) adheres to the usual form (see Eqs.~\eqref{eq:2loop:CLs} and \eqref{eq:3loop:CLs}) with the alternating sign:
\begin{align}
 \S_{4,\cl}^{\JA}(\rho) = +\asb^4\,\frac{L^4}{4!}\,\CFfour\,\F_4^{\JA},
\end{align}
where the four loop CLs coefficient for the \(k_t\) clustering is given by \cite{Khelifa-Kerfa:2024hwx, Khelifa-Kerfa:2024hwz}:
\begin{align}\label{eq:4loop:F4-KT}
\F_4^{\ktt} &= \Biggl[
	\int_{1_\out}\!\int_{2_\out}\!\int_{3_\out}\!\int_{4_\inn} \O_{14}\,\O_{24}\,\O_{34}\notag\\
	&\quad+\int_{1_\out}\!\int_{2_\out}\!\int_{3_\inn}\!\int_{4_\inn}\!\Bigl[\O_{13}\,\O_{24}(-1+\Ob_{14}\,\Ob_{34})\notag\\
	&\qquad+\O_{23}\bigl(-\O_{13}-\O_{14}+\Ob_{24}\,\Ob_{34}(1-\Ob_{13}\,\Ob_{14})\bigr)\Bigr]\notag\\
	&\quad+\int_{1_\out}\!\int_{2_\inn}\!\int_{3_\out}\!\int_{4_\inn}\O_{12}\,\O_{34}(-1+\Ob_{14}\,\Ob_{24})\notag\\
	&\quad+ \int_{1_\out}\!\int_{2_\inn}\!\int_{3_\inn}\!\int_{4_\inn}\O_{12}(-1+\Ob_{13}\,\Ob_{23})\notag\\
	&\qquad\times(-1+\Ob_{14}\,\Ob_{24}\,\Ob_{34})
\Biggr]\,w_{ab}^1\,w_{ab}^2\,w_{ab}^3\,w_{ab}^4
= 3.358.
\end{align}
This approximates to \(31\,\zeta_4/10\).
The C/A CLs coefficient, \(\F_4^{\CAA}\), is the sum of the \(k_t\) coefficient and an additional term \(f_4^{\CAA}\), which receives contributions from the same four regions of phase space but with different clustering functions. The explicit form of \(f_4^{\CAA}\) is lengthy and is given in Ref.~\cite{CAclus}. Its integrals may be performed using the \texttt{Cuba} library. The final result is:
\begin{align}\label{eq:4loop:F4-CA}
 \F_4^{\CAA} = \F_4^{\ktt} + f_4^{\CAA} = 3.358 - 0.444 = 2.913.
\end{align}
Note that, unlike the three loop C/A coefficient \(f_3^{\CAA}\), which shares the sign of \(\F_3^{\ktt}\) and hence adds, the four loop term \(f_4^{\CAA}\) is of opposite sign and thus subtracts. Consequently, \(\F_4^{\CAA}\) is smaller than \(\F_4^{\ktt}\). Moreover, \(\F_4^{\ktt}/4! = 0.44\) and \(\F_4^{\CAA}/4! = 0.29\), both of which are smaller than their three and two loop counterparts, ensuring convergence of the perturbative series.

The four loop NGLs contribution, \(\S_{4,\ng}^{\JA}\), for a given jet algorithm (JA) is given, as at two and three loops (Eqs.~\eqref{eq:2loop:NGLs} and \eqref{eq:3loop:NGLs}), by:
\begin{align}\label{eq:4loop:NGLs}
 \S_{4,\ng}^{\JA}(\rho) &= -\asb^4\,\frac{L^4}{4!}\Bigl[-\CFcub\,\CA\,\G_{4,a}^{\JA}
   - \CFsq\,\CAsq\,\G_{4,b}^{\JA} \notag\\
&\qquad\quad + \CF\,\CAcub\,\G_{4,c}^{\JA}
   + \CF\,\CAsq\bigl(\CA - 2\CF\bigr)\,\G_{4,d}^{\JA}\Bigr],
\end{align}
where, as previously stated, clustering affects all colour channels. For the anti-\(k_t\) algorithm, only \(\G_{4,c}^{\akt}=29\zeta_4/8=3.92\) and \(\G_{4,d}^{\akt}=-\zeta_4/2=-0.54\) contribute.  For \(k_t\) clustering, numerical integrations yield:
\begin{align}\label{eq:4loop:G4-KT-parts}
 \G_{4,a}^{\ktt} = 2.81,\quad
 \G_{4,b}^{\ktt} = 0.27,\quad
 \G_{4,c}^{\ktt} = 1.00,\quad
 \G_{4,d}^{\ktt} = -0.73,
\end{align}
where, following Ref.~\cite{Khelifa-Kerfa:2024hwx}, a minus sign has been extracted from \(\G_{4,c}^{\ktt}\) into the front of the colour factor \(\CF\,\CAcub\).

The C/A coefficients are, as usual, the sum of the \(k_t\) coefficients and additional terms from regions not covered by the \(k_t\) algorithm:
\begin{align}\label{eq:4loop:g4-CA}
 g_{4,a}^{\CAA} = -1.26,\;
 g_{4,b}^{\CAA} = 1.25,\;
 g_{4,c}^{\CAA} = 0.58,\;
 g_{4,d}^{\CAA} = 0.21.
\end{align}
The interested reader is referred to Ref.~\cite{CAclus} for further details. The final C/A NGLs coefficients are then:
\begin{align}\label{eq:4loop:G4-CA-parts}
 \G_{4,a}^{\CAA} = 1.56,\;
 \G_{4,b}^{\CAA} = 1.53,\;
 \G_{4,c}^{\CAA} = 0.42,\;
 \G_{4,d}^{\CAA} = -0.53.
\end{align}
To quantify the impact of \(k_t\) and C/A clusterings on the hemisphere mass distribution at this order we combine CLs and NGLs and compare across the three algorithms:
\begin{align}
& -\frac{1}{\asb^4 L^4} \S_{4,\ng}^{\akt} = -5.79,\notag\\
& \frac{1}{\asb^4 L^4}\bigl[\S_{4,\cl}^{\ktt}-\S_{4,\ng}^{\ktt}\bigr] = 0.08,\notag\\
& \frac{1}{\asb^4 L^4}\bigl[\S_{4,\cl}^{\CAA}-\S_{4,\ng}^{\CAA}\bigr] = 1.32.
\end{align}
For both clusterings there is a significant reduction in the non‐global logarithmic contribution—up to $99 \%$ for \(k_t\) and $77 \%$ for C/A clustering.

Since the eikonal amplitudes squared have not been fully determined beyond four loops, we are unable to carry out calculations of the hemisphere mass distribution beyond four loops. The form of the distribution at two, three and four loops (Eqs.~\eqref{eq:2loop:Sigma-FinalForm}, \eqref{eq:3loop:Sigma-FinalForm} and \eqref{eq:4loop:Sigma-FinalForm}) suggests a pattern of exponentiation, not only for the primary independent emissions—already known to exponentiate into the Sudakov form factor—but also for the CLs and NGLs contributions. It has not, however, been possible to resum them analytically into a closed form. Including full colour dependence, the present work marks the state of the art in fixed‐order computations of NGLs and CLs for all three jet algorithms. For the anti-$k_t$ algorithm, and using the large-$N_c$ approximation, fixed-order calculations have been extended up to twelve loops \cite{Caron-Huot:private} for the hemisphere mass observable in $e^+e^-$. As mentioned in the introduction, numerical resummations of CLs and NGLs exist only for the anti-$k_t$ and $k_t$ algorithms, and full‐colour resummation is available only for the anti-$k_t$ case and only to SL accuracy.

In the next section we address the resummation of all three types of large logarithms present in the hemisphere mass distribution, namely global (primary emissions), CLs and NGLs.

\section{All-orders and comparison to MC}
\label{sec:Resum}

The general form of the resummation of large logarithms that occur in the distribution of non-global observables may be cast as \cite{Dasgupta:2001sh, Banfi:2010pa}:
\begin{align}\label{eq:Resum:Sigma-tot}
 \S(\rho) = \S_{\mathrm{P}}(\rho)\,\S_{\ng}^{\JA}(\rho)\,\S_{\cl}^{\JA}(\rho),
\end{align}
where \(\S_{\mathrm{P}}\), \(\S_{\ng}^{\JA}\) and \(\S_{\cl}^{\JA}\) represent the global (Sudakov), NGLs and CLs form factors, respectively. The global Sudakov form factor resums soft-collinear and hard-collinear independent emissions, including the running of the coupling, and can be written up to NLL as \cite{Catani:1992ua, Banfi:2010pa}:
\begin{align}\label{eq:Resum:Global}
 \S_{\mathrm{P}}(\rho) = \frac{\exp\bigl[-\cR - \gamma_E\,\cR'\bigr]}{\Gamma\bigl(1 + \cR'\bigr)},
\end{align}
where \(\Gamma\) is the gamma function and the NLL radiator \(\cR\) and its derivative \(\cR'\) are
\begin{align}\label{eq:Resum:GlobalRadiator}
 \cR &= -L\,g_1(\lambda) - g_2(\lambda),
 &
 \cR' &= -\frac{\p}{\p L}\bigl[L\,g_1(\lambda)\bigr],
\end{align}
with \(L = \ln(1/\rho)\) and \(\lambda = \beta_0\,\as(Q^2)\,L\). The LL and NLL functions \(g_1\) and \(g_2\) read:
\begin{subequations}\label{eq:Resum:g1-g2-Funcs}
\begin{align}
 g_1(\lambda) &= -\frac{\CF}{2\pi\,\beta_0\,\lambda}\Bigl[(1-2\lambda)\ln(1-2\lambda) \notag\\
             &\quad -2(1-\lambda)\ln(1-\lambda)\Bigr],
\\
 g_2(\lambda) &= -\frac{\CF\,K}{4\pi^2\,\beta_0^2}\bigl[2\ln(1-\lambda)-\ln(1-2\lambda)\bigr]\notag\\
 &\quad- \frac{\CF\,\gamma_E}{\pi\,\beta_0}\bigl[\ln(1-\lambda)-\ln(1-2\lambda)\bigr]\notag\\
 &\quad - \frac{\CF\,\beta_1}{2\pi\,\beta_0^3}\Bigl[\ln(1-2\lambda)-2\ln(1-\lambda)
 +\tfrac12\ln^2(1-2\lambda) \notag\\
&\quad-\ln^2(1-\lambda)\Bigr] - \frac{3\,\CF}{4\pi\,\beta_0}\ln(1-\lambda).
\end{align}
\end{subequations}
The two-loop running coupling in the \(\overline{\mathrm{MS}}\) scheme is \cite{Catani:1992ua}
\begin{align}
 \as(Q^2) = \frac{\as(\mu^2)}{1 - v}\Bigl[1 - \as(\mu^2)\,\frac{\beta_1}{\beta_0}\,\frac{\ln(1-v)}{1-v}\Bigr],
\end{align}
where \(v=\as(\mu^2)\,\beta_0\,\ln(\mu^2/Q^2)\), and the beta-function coefficients are
\begin{align}
 \beta_0 = \frac{11\,\CF - 2\,n_f}{12\pi},
 \quad
 \beta_1 = \frac{17\,\CAsq - 5\,n_f\,\CA - 3\,n_f\,\CF}{24\pi^2},
\end{align}
with \(n_f=5\). The constant \(K\) \cite{Catani:1990rr} is
\begin{align}
 K = \CA\Bigl(\frac{67}{18} - \frac{\pi^2}{6}\Bigr) - \frac{5}{9}\,n_f.
\end{align}

The NGLs form factor, \(\S_{\ng}^{\JA}(\rho)\), may be parametrised, for the anti-\(k_t\) algorithm, by the following formula, first introduced in the original paper on NGLs \cite{Dasgupta:2001sh}:
\begin{align}\label{eq:Resum:NGLs-akt}
 \S_{\ng}^{\akt}(\rho) = \exp\bigl[ -\CF\,\CA\,\G_2^{\akt}\,\frac{1 + (0.86\,\CA\,t)^2}{1 + (0.86\,\CA\,t)^{1.33}}\,t^2  \bigr],
\end{align}
where \(\G_2^{\akt} = \zeta_2\) and the SL evolution variable \(t\) is defined by
\begin{align}\label{eq:Resum:t-def}
 t = \frac{1}{2\pi}\int_{\rho}^{1}\frac{\d x}{x}\,\as(xQ)
     = -\frac{1}{4\pi\beta_0}\ln\bigl[1 - 2\,\beta_0\,\as\,\ln(1/\rho)\bigr].
\end{align}
The second equality is the one-loop expansion, sufficient for our SL accuracy. Note that at fixed order \(t=\as\,L/2\pi=\asb\,L/2\). It was shown in Refs.~\cite{Dasgupta:2001sh,Khelifa-Kerfa:2015mma} that the all-order resummed form factor is well approximated by the exponentiation of the two loop result over a wide range of \(t\), not only for \(e^+e^-\) processes but also in hadron–hadron collisions \cite{Khelifa-Kerfa:2024udm}.

Since the MC of Ref.~\cite{Dasgupta:2001sh} does not provide results for the hemisphere mass distribution, nor do the numerical solutions of the BMS equation \cite{Banfi:2002hw}, for the \(k_t\) or C/A algorithms, we shall approximate the all-order resummation by the exponential of the fixed-order result up to four loops. To this end we write:
\begin{align}\label{eq:Resum:NGLs-kt_CA}
 \S_{\ng}^{\JA}(\rho) = \exp\Biggl[ -\sum_{n=2}^{4}\frac{(-1)^n}{n!}\,\tilde{\G}_n^{\JA}\,(2t)^n \Biggr],
\end{align}
where “JA” refers to the anti-\(k_t\), \(k_t\) and C/A algorithms, and we have replaced \(\as L\) by \(2t\) as explained above. The NGLs coefficient \(\tilde{\G}_n^{\JA}\) includes the corresponding colour factors; e.g.\ \(\tilde{\G}_2^{\JA}=\CF\CA\,\G_2^{\JA}\), \(\tilde{\G}_3^{\JA}=\CFsq\CA\,\G_{3,a}^{\JA}+\CF\CAsq\,\G_{3,b}^{\JA}\), and similarly for higher-loop coefficients.

Similarly, we approximate the all-order CLs form factor by the exponential of the fixed-order results:
\begin{align}\label{eq:Resum:CLs-kt_CA}
 \S_{\cl}^{\JA}(\rho) = \exp\Biggl[\sum_{n=2}^4 \frac{(-1)^n}{n!}\,\CF^{\!\!\!\! n}\,\F_n^{\JA}\,(2t)^n\Biggr],
\end{align}
where “JA” refers only to the $k_t$ and C/A algorithms, since there are no CLs coefficients for the anti-$k_t$ algorithm.

\begin{figure}[th]
\centering
\includegraphics[scale=0.6]{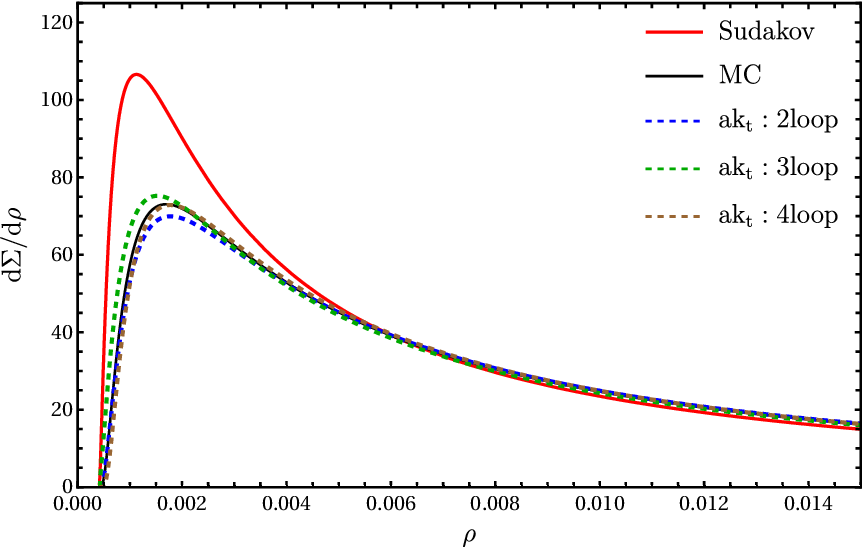}
\caption{Resummed differential distribution of the hemisphere mass for the anti-$k_t$ algorithm with various approximations of the NGLs form factor.}
\label{fig:MC-akt}
\end{figure}

In Fig.~\ref{fig:MC-akt} we plot the resummed differential distribution, $\d\S/\d\rho$, for the anti-$k_t$ algorithm without NGLs (i.e.\ including only the Sudakov form factor) and with their inclusion. For the latter we show both the MC-parametrised form \eqref{eq:Resum:NGLs-akt} and the exponential of the fixed-order results \eqref{eq:Resum:NGLs-kt_CA}, where “2loop” includes only the $n=2$ term in the sum of Eq.~\eqref{eq:Resum:NGLs-kt_CA}, “3loop” includes $n=2,3$, and so on. The “4loop” curve (brown) approximates the MC distribution better than the two and three loop curves. The MC and 4loop curves exhibit a shift in the peak position and a reduction in its height by about $31\%$.

\begin{figure}[th]
\centering
\includegraphics[scale=0.6]{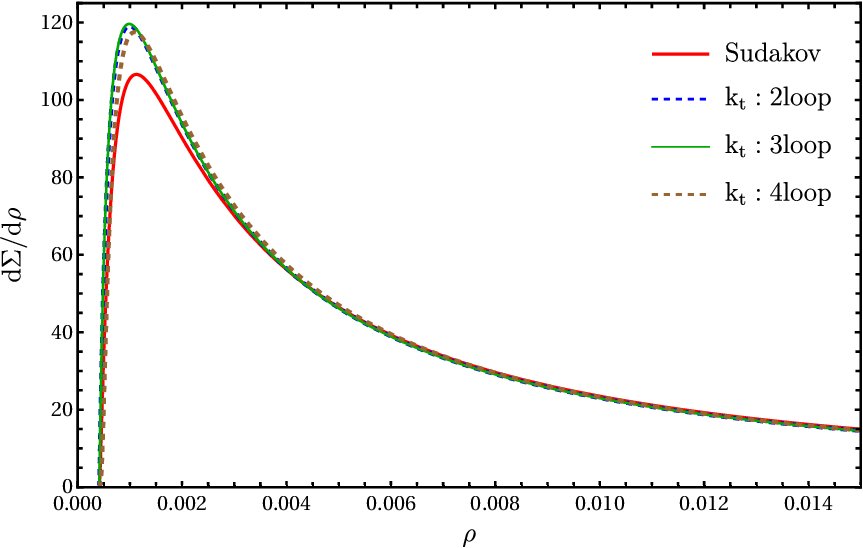}\vskip0.5em
\includegraphics[scale=0.6]{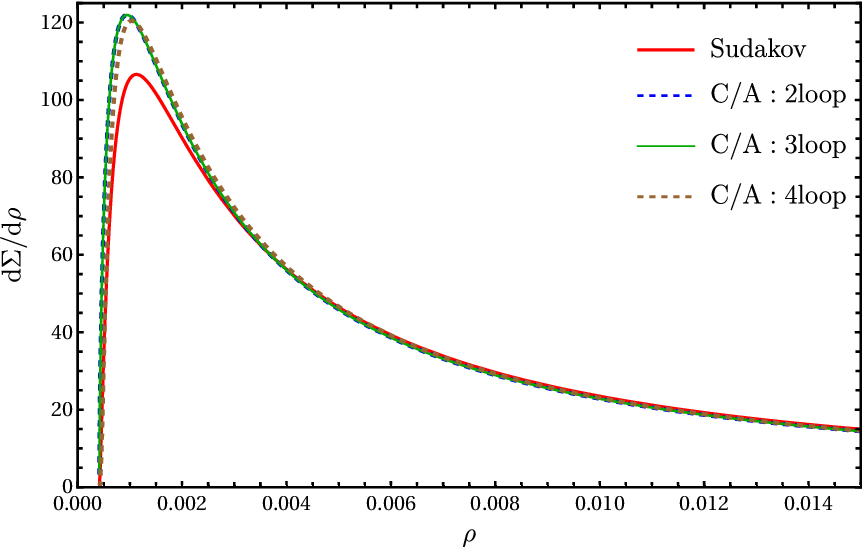}
\caption{Resummed differential distribution of the hemisphere mass for the $k_t$ (top) and C/A (bottom) algorithms with various approximations of the CLs and NGLs form factors.}
\label{fig:Resum-kt-ca}
\end{figure}

Similarly, in Fig.~\ref{fig:Resum-kt-ca} we plot the resummed differential distributions for the $k_t$ and C/A algorithms, where the CLs and NGLs are represented by the exponentials of the fixed-order results (Eqs.~\eqref{eq:Resum:CLs-kt_CA} and \eqref{eq:Resum:NGLs-kt_CA}).
Unlike the anti-$k_t$ case, the $k_t$ and C/A algorithms induce an increase in the Sudakov peak of about $10\%$–$12\%$ for $k_t$ and $12\%$–$14\%$ for C/A, with the “4loop” curve showing the smallest increase. Since the largest contribution arises from the two loop CLs coefficient—more than six times larger than that of the jet-mass observable \cite{Banfi:2010pa, Khelifa-Kerfa:2011quw, Delenda:2012mm, Khelifa-Kerfa:2024hwx}—the higher-loop contributions merely induce fractional changes. Indeed, combining CLs and NGLs at two loops yields $+0.68 \,(\asb L)^2$ for the hemisphere mass versus $-1.3\,(\asb L)^2$ for the jet mass. Thus, the increase around the peak is driven predominantly by the large two loop CLs result (see Sec.~\ref{sec:2loop}).

\begin{figure}[th]
\centering
\includegraphics[scale=0.6]{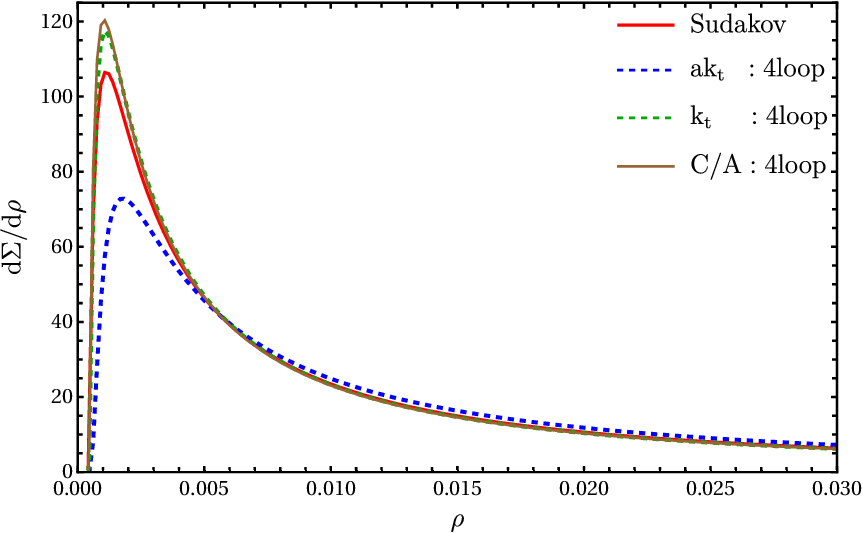}
\caption{Resummed differential distribution of the hemisphere mass for all three jet algorithms.}
\label{fig:Resum-JetAlgs}
\end{figure}

Finally, in Fig.~\ref{fig:Resum-JetAlgs} we compare the resummed distributions for all three algorithms, using the exponential approximations up to four loops.
Unlike the anti‐$k_t$ algorithm, the $k_t$ and C/A distributions induce no shift in the position of the Sudakov peak but increase its height by approximately $11\%$ and $14\%$, respectively. Moreover, both the $k_t$ and C/A algorithms better reproduce the tail of the Sudakov distribution. By contrast, the anti‐$k_t$ distribution exceeds the pure Sudakov starting at $\rho\sim0.006$.

In the case of the jet mass observable, the $k_t$ algorithm produces a distribution that nearly coincides with the Sudakov form factor \cite{Khelifa-Kerfa:2024hwx}, suggesting that one might neglect non‐global effects altogether. The hemisphere mass observable, however, does not admit such an “optimal” algorithm: all three jet algorithms yield distributions that differ significantly from the pure Sudakov result. Consequently, the non‐global nature of the hemisphere mass must carefully be considered in any precise phenomenological analysis.

\section{Conclusions}
\label{sec:Conclusion}

In this work, we have presented the first four loop computation of the hemisphere mass
distribution in $e^+ e^-$ annihilation, incorporating the effects of sequential recombination jet
algorithms --specifically, the $k_t$ and C/A variants. By
employing eikonal methods and strong-energy ordering, we have systematically derived, at leading logarithmic accuracy and with full colour,
the complete structure of both abelian (clustering-induced) and non-abelian non-global
logarithms at this perturbative order. Our results provide crucial insights into the
interplay between jet algorithms and non-global QCD effects, with significant implications
for precision phenomenology.

The two loop $k_t$ (and, consequently, C/A) calculations yield larger CLs coefficients and smaller NGLs coefficients compared to the jet‐mass observable. This may result from the larger phase space available for independent emissions (the origin of CLs), particularly in collinear regions near the outgoing hard partons, together with a more restricted phase space for NGLs. These algorithms crucially redistribute soft gluons across the inter‐hemisphere boundary through drag-in and drag-out mechanisms. As the C/A algorithm is energy-independent, it permits additional distance‐orderings between gluons and thus accesses a larger phase space (encompassing the $k_t$ configurations). Consequently, larger coefficients are expected for both CLs and NGLs in C/A.

Three and four loop calculations follow the same procedure as at two loops. While the CLs and NGLs coefficients grow with loop order, they are suppressed by the factors $3!$ and $4!$ and by additional powers of $\CF$, thereby ensuring the convergence of the perturbative series. Moreover, CLs and NGLs generally exhibit opposite signs, producing a partial cancellation and yielding smaller net corrections to the Sudakov distribution than in the anti-$k_t$ case.

Our analytical calculations up to four loops confirm that the hemisphere mass distribution exhibits an exponentiation pattern for both primary and secondary emissions. This observation suggests an all-orders form for CLs and NGLs as exponentials of the fixed‐order results. Comparison with anti-$k_t$ Monte Carlo data shows that retaining all terms up to four loops in the exponent provides an excellent approximation.

In the absence of numerical resummation for $k_t$ and C/A, we compare their resummed distributions with the pure Sudakov form factor. Whereas anti-$k_t$ reduces the distribution peak, shifts it to larger masses, and overestimates the tail, the $k_t$ and C/A algorithms increase the peak (offsetting nearly half of the anti-$k_t$ reduction), maintain its position, and closely reproduce the tail. This contrasts with the jet mass observable, for which the $k_t$ algorithm coincides almost exactly with the Sudakov form factor—rendering non-global logarithms negligible. For the hemisphere mass observable, however, precise predictions require the full inclusion of both abelian (clustering) and non-abelian non-global logarithms.

It is essential that parton‐shower algorithms incorporate both clustering and non‐global logarithmic corrections to reproduce the perturbative structure of non‐global observables accurately. The methodology presented, based on strong‐energy ordering and eikonal amplitudes, is directly applicable to a wide class of non‐global observables in both \(e^+e^-\) and hadron–hadron collisions. Furthermore, the same strategy can be applied to any member of the generalised \(k_t\) family, enabling systematic calculations of non‐global effects across different jet algorithms.

\section*{Acknowledgements}

The researchers extend their sincere gratitude to the Deanship of Scientific Research at the
Islamic University of Madinah for the support provided to the Post-Publishing Program.


\bibliographystyle{spphys}
\bibliography{Refs}

%
%
%
%

\end{document}